\documentclass[aps,prl,twocolumn,showpacs,groupedaddress]{revtex4}  
\usepackage{graphicx}  
\usepackage{dcolumn}   
\usepackage{bm}        
\usepackage{amssymb}   
\usepackage{subfigure}
\usepackage{amsmath,amsfonts}
\usepackage{setspace}

\begin{document}
\title{Two phase transitions in the two-dimensional nematic 3-vector model with no quasi long-range order: Monte Carlo simulation of the density of states}

\author{B. Kamala Latha$^{1}$}
\author{V.S.S. Sastry$^{2}$}

\affiliation {$^{1}$School of Physics,University of Hyderabad, Hyderabad 500046, India}

\affiliation{$^{2}$Centre for Modelling, Simulation and Design, University of 
Hyderabad, Hyderabad 500046, India}

\date{\today}

\begin{abstract}
The presence of stable topological defects in a two-dimensional 
(\textit{d} = 2) liquid crystal model allowing molecular reorientations 
in three dimensions (\textit{n} = 3) was largely believed to induce 
defect-mediated Berzenskii-Kosterlitz-Thouless (BKT) type transition to a
low temperature phase with quasi long-range order. However, 
earlier Monte Carlo (MC) simulations could not establish certain essential 
signatures of the transition, suggesting further investigations. We study 
this model by computing its equilibrium properties through MC simulations,
based on the determination of the density of states of the system. Our 
results show that, on cooling, the high temperature disordered phase 
deviates from its initial progression towards the topological transition, 
crossing over to a new fixed point, condensing into a nematic phase 
with exponential correlations of its director fluctuations. The 
thermally induced topological kinetic processes continue, however limited 
to the length scales set by the nematic director fluctuations, and lead 
to a second topological transition at a lower temperature. We argue that 
in the (\textit{d} = 2, \textit{n} = 3) system with a biquadratic Hamiltonian, the presence of 
additional molecular degree of freedom and local $Z_{2}$ symmetry associated with lattice sites, 
together promote the onset of an additional relevant scaling field at matching length scales in the high temperature region, leading to a crossover.

\end{abstract}

\maketitle

Two-dimensional (\textit{d} = 2) liquid crystal (LC) models with
molecular reorientations in three dimensions (\textit{n} = 3) host
stable topological point defects (disclination points) with half
integral charge \cite{Mermin} owing to their apolar order parameter
(OP) geometry (real two-dimensional projective space $RP^{2}$), and 
are predicted to undergo a topological phase transition \cite{Kunz}.
Several Monte Carlo (MC) studies on Lebwohl-Lasher (LL) model \cite{LL}
confined to a two-dimensional square lattice were carried out based on
the Metropolis algorithm, alluding to a Berenzskii-Kosterlitz-Thouless
type \cite{BKT} topological transition to a low-temperature phase with
quasi long-range order (QLRO) \cite{CZ,Kunz,BP,Mondal, Dutta, Shabnam}.
This assignment could not be conclusive however, primarily because 
of the absence of size-invariant Binder's cumulant \cite{Binder} 
at the reported transition temperature \cite{Paredes, Sanchez}. Recent 
comparative analysis of the MC data based on finite size scaling 
criterion \cite{Tomita} distinguishes the $RP^{2}$-systems from other
two-dimensional magnetic systems, \textit{viz}. 2d-XY and 
2d-Heisenberg models, and concludes that the 2d LC systems could only
have pseudo-critical regions. A plausible conjecture advanced to
account for the observed inconsistency has been the presence of an
underlying subtle and persistent crossover in the model \cite{Shabnam,
Paredes, Sanchez}. 

In this context, we examined this model with a different MC sampling
procedure: we derived the equilibrium properties of this model by
first computing its density of states (DoS) and then constructing
equilibrium ensembles. The DoS was obtained with a variant of the MC
procedure, - entropic sampling technique \cite{WL} -, which is geared
to access the configuration space uniformly with respect to energy.  
We used a modified version of the Wang-Landau algorithm
\cite{jayasri05} augmented with \textit{frontier-sampling} \cite{Zhou}
technique to enhance its efficacy \cite{BKL15}. We constructed entropic
ensembles comprising of microstates distributed uniformly with energy
by performing a random walk biased by the DoS of the system. We 
distinguish the equilibrium ensembles obtained by reweighting procedure
from the entropic ensemble (say, \textit{RW}-ensembles) from those
obtained based on the Metropolis algorithm, 
(\textit{B}-ensembles) \cite{BKL15}, while comparing their equilibrium
averages of different physical properties.     
 
We considered a square (\textit{d} = 2) lattice of variable size 
$L\times L$ (\textit{L} = 50, 80, 100, 120, 150), each lattice site
 hosting a manifold of directions in three dimensional space
  (\textit{n} = 3). The interactions are described by the 
  Lebwohl-Lasher Hamiltonian, 
$H= -\epsilon\sum_{<i,j>}  P_{2} (\cos \theta_{ij})$, with the 
summation covering all the nearest neighbours, and prescribing periodic 
boundary conditions. $P_{2}$ is the second Legendre polynomial and 
$\theta_{ij}$ is the angle between the neighbouring molecules. The 
temperature ($T$) is reported in reduced units scaled by the coupling 
strength $\epsilon$. The \textit{RW}-ensembles are constructed from the 
entropic ensemble \cite{BKL15}, in the temperature range of interest 
(\textit{T} = 0.4 to 1.0, with a resolution of 0.001). 
We computed, as a function of \textit{T}, the averages of energy
\textit{E}, nematic order parameter \textit{S}, nematic
susceptibility \textit{$\chi$}, as well as Binder's cumulant
\textit{$R_{4}$} associated with \textit{S} to monitor the
transition (\cite{BKL15}). In addition, we also calculated the average
values of the density of unbound defects $\textit{D}_\textit{t}$ as 
well as topological order parameter $\delta$ based on algorithms 
described in \cite {Kunz, Dutta}. Orientational pair correlation 
function $G(r_{ij}) = <P_{2} (\cos \theta_{ij})> $ was computed (at 
\textit{L} = 150) at about 40 temperatures covering the above range.
 The averages of \textit{E}, \textit{S}, $\textit{D}_\textit{t}$ 
and $\delta$ have statistical errors (estimated using Jack-knife
algorithm \cite{BKL}) typically of the order of 1 in $10^{4}$, while
the higher moments ($C_{v}$, \textit{$\chi$}, \textit{$R_{4}$}) are
 relatively less accurate, estimated to be about 5 in $10^{3}$. 

\begin{figure}
\centering
\includegraphics[width=0.43\textwidth]{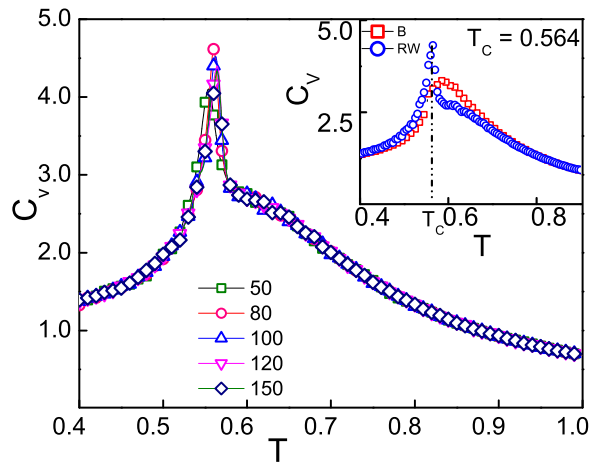}
\caption{(color online) Temperature variation of specific heat 
(per site)from \textit{RW}-ensembles at sizes \textit{L} = 50, 
80, 100, 120, 150. Inset shows qualitatively different temperature variations of $C_{v}$ from \textit{RW}- and \textit{B}-ensembles
 at \textit{L} = 150.} 
\label{fig:1}
\end{figure}
We now present qualitatively differing features of the physical
parameters obtained from the two types of sampling procedures, leading
to a discussion on the interpretation of our observations. Fig.~\ref{fig:1} 
shows an essentially size-independent temperature variation of $C_{v}$ 
(per site) as obtained from \textit{RW}-ensembles at different system 
sizes, indicating an initial development of a broad cusp on cooling, but 
yielding to an abrupt sharp peak located at 0.564 (\textit{L}=150). This 
is to be contrasted with the broad cusp obtained from the \textit{B}-ensemble 
at the same size (shown in the Inset of Fig.~\ref{fig:1}), which is in 
accord with the data reported earlier.  
\begin{figure}
\centering
\includegraphics[width=0.43\textwidth]{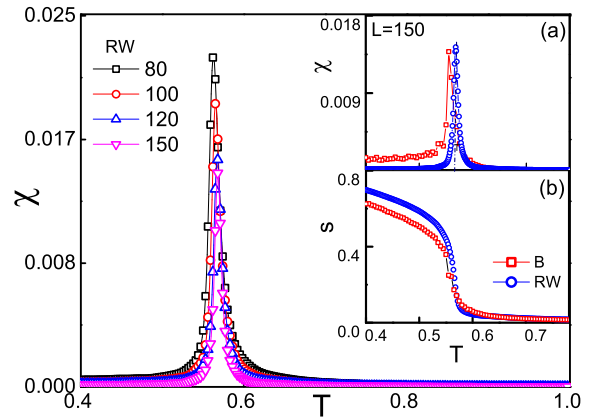}
\caption{(color online) Temperature variation of nematic susceptibility $\chi$ 
from \textit{RW}-ensembles at sizes \textit{L} = 80, 100, 120 and 150. Insets (a) and (b)
depict the comparison of the temperature variations of $\chi$ and nematic order
 \textit{S} from \textit{RW}-ensembles with \textit{B}-ensembles at \textit{L} = 150.} 
\label{fig:2}
\end{figure}
Fig.~\ref{fig:2} depicts the temperature dependence of 
$\chi$ from \textit{RW}-ensembles as a function of size. Its temperature 
variations as obtained from the two ensembles are compared in the 
Inset (a) at \textit{L} = 150. The corresponding order parameters are 
shown in Inset (b). The values of the order parameter \textit{S} in the 
low temperature phase were found to decrease with size as computed 
from \textit{B}-ensembles (consistent with the QLRO regime), while 
\textit{RW}-ensembles essentially report its size-independence for 
$L\geq 80$ (not shown here). Low temperature values of the susceptibility 
also qualitatively differ (Inset (a)): it is non-zero and diverging with 
size in the \textit{B}-ensembles , while its value quickly tends to zero 
with the present sampling procedure. Also, the $\chi$ peaks in the present 
study shift slightly towards higher temperature with size, very similar 
to $C_{v}$.

\begin{figure}
\centering
\includegraphics[width=0.46\textwidth]{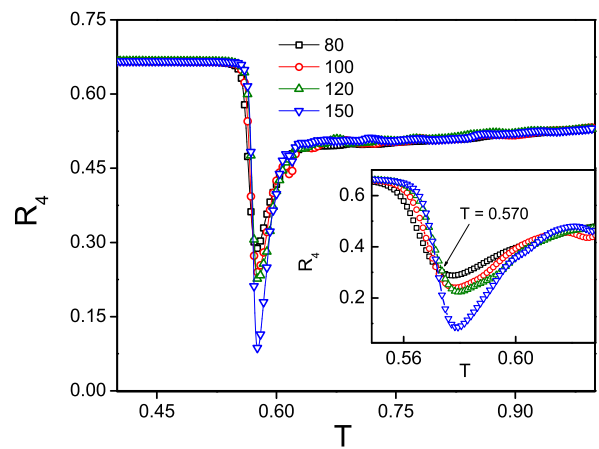}
\caption{(color online) Temperature variation of Binder's cumulant of
the nematic order \textit{$R_{4}$} from \textit{RW}-ensembles at sizes 
\textit{L} = 80, 100, 120 and 150. Inset shows the magnified version near the transition.} 
\label{fig:3}
\end{figure}
The absence, from the earlier Monte Carlo studies, of a size-invariant 
Binder's cumulant (\textit{$R_{4}$}) at the predicted transition temperature 
has been a major obstacle to unambiguously assign the transition as 
defect-mediated, required to explain the observed low temperature QLRO 
phase \cite{Paredes, Sanchez}. From our data based on the DoS, a 
size-independent cumulant value was obtained at $T$ = 0.570 $\pm$ 0.001 
(Fig.~\ref{fig:3} and Inset), providing a confirmatory evidence of a 
(continuous) transition at this temperature, as also representing the 
thermodynamic limit of the size-dependent $C_{v}$ peak positions 
Fig.~\ref{fig:1}. 
\begin{figure}
\centering
\includegraphics[width=0.55\textwidth]{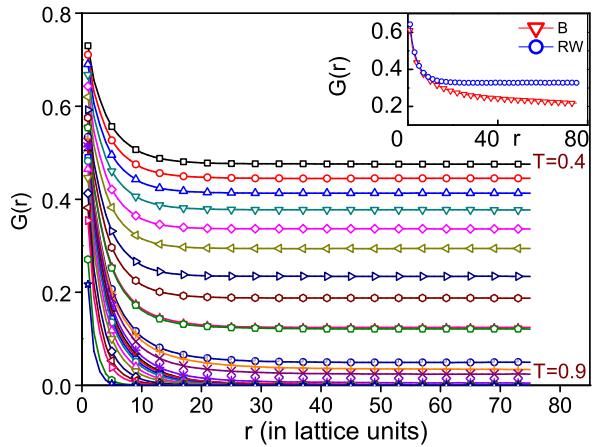}
\caption{(color online) Variation of the orientational pair 
correlation function $G(r)$ with lattice distances 
(\textit{L} = 150) in the temperature range \textit{T} = 0.4 
to 0.9. The inset compares the $G(r)$ obtained from
\textit{B}- and \textit{RW}-ensembles in the low 
temperature phase, at \textit{T} = 0.5.} 
\label{fig:4}
\end{figure}
We investigate the nature of the low temperature phase by computing the 
spatial dependences of the orientational correlations of LC molecules at 
\textit{L} = 150. Variations of $G(r)$ with distance (in lattice units), 
at different temperatures spanning the window \textit{T} = 0.4 to 0.9, are 
shown in Fig.~\ref{fig:4}. Qualitatively differing from the earlier 
observations of power law variation at low temperatures, $G(r)$  fit 
very well to exponential decays, leading to the determination of the 
correlation length $\xi(T)$ (to within about 1$\%$ error). The Inset 
compares $G(r)$ obtained from \textit{RW}- and \textit{B}-ensembles 
at \textit{T} = 0.5 (low-temperature phase), at \textit{L} = 150. 
The \textit{B}-ensemble data require a power law for a satisfactory fit, 
as concluded by the earlier studies. Presence of a single length scale 
in the system at any temperature is the major indicator 
signalling a qualitative departure from the current interpretation 
of the low temperature phase in terms of QLRO.

\begin{figure}
\centering
\includegraphics[width=0.43\textwidth]{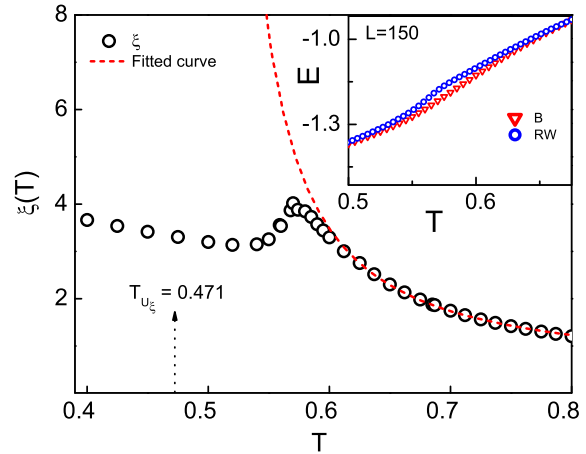}
\caption{(color online) Variation of correlation length $\xi$  as a 
function of temperature. The dotted line represents the extrapolated 
divergence of $\xi$ based on high temperature data (see text). Inset shows the variation of energy with temperature near the peak position from both the ensembles.}
\label{fig:5}
\end{figure}
 We show the temperature variation of $\xi(T)$ in Fig.~\ref{fig:5}. 
Starting from the high temperature side, $\xi(T)$ initially tends to 
diverge, but deviates away near $\textit{T}\approx 0.6$ to form a cusp at 
0.564. We analyse the high temperature data ($\xi(T)$ \textit{vs} \textit{T} 
in Fig.~\ref{fig:5}) assigning its divergence as due to the temperature 
dependent kinetics of the unbound defects, given by 
$\xi(T) \sim \exp[\frac{A}{(T - T_{U})^{1/2}}]$
in the mean-field limit \cite{BKT}. Here $T_{U}$ is the limiting 
temperature for the unbound defects to exist and A is a constant. The 
present data fit to this expression very satisfactorily till about 
$\textit{T} \simeq 0.6$ (Fig.\ref{fig:5}), yielding an estimate of 
the unbinding transition temperature $T_{U_{\xi}}$ = 0.471 $\pm$ 0.005. 
This implies that, but for the interruption by the transition at $T_{c}$, 
the system would have proceeded to a direct topological transition with 
a broad $C_{v}$ cusp terminating its critical contribution in the 
neighbourhood of  $T_{U_{\xi}}$ as a weak essential singularity. The 
departure of the observed $\xi$ from its expected divergence (shown as 
dotted lines in Fig.~\ref{fig:5}) forming a cusp at \textit{T} = 0.564, 
as well as concomitant development of a sharp $C_{v}$ peak at the same 
temperature show a crossover of the system towards a new fixed point 
from its initial progression towards a topological transition. The 
inset of this figure shows the differences in the energies (per site) as 
computed by the two MC procedures in the crossover region. The Metropolis 
algorithm accesses lower energy microstates in locating the regions of  
equilibrium ensembles, while the density of states computation samples 
relatively higher energy microstates, - both being guided however by 
the same requirement of free energy minimization. 
  
\begin{figure}
\centering
\includegraphics[width=0.5\textwidth]{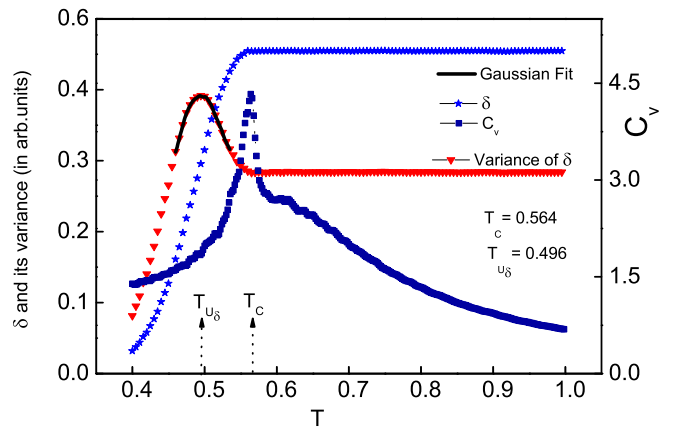}
\caption{(color online) Temperature variation of topological order 
$\delta$ (stars)  and of its variance (triangles) superposed
on the specific heat $C_{v}$ profile (squares). Solid line indicates a
local Gaussian fit to the cusp of the variance.} 
\label{fig:6}
\end{figure}
\begin{figure}
\centering
\includegraphics[width=0.5\textwidth]{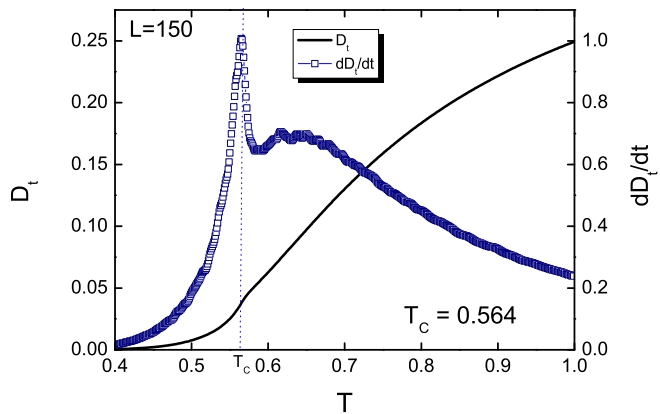}
\caption{(color online) Temperature variation of defect density
 $D_{t}$ from \textit{RW}-ensembles at \textit{L} = 150, superposed by its temperature derivative.} 
\label{fig:7}
\end{figure}
The temperature variation of $\delta$ associated with the topological 
order \cite{Kunz}, along with $C_{v}$ is depicted in  Fig.~\ref{fig:6}. 
The variance of $\delta$ (in arbitrary units for plotting convenience) 
is also shown in the same figure. The location of the peak of this 
variance is determined by fitting a local Gaussian, leading to an 
estimate of the unbinding transition temperature at  
$T_{U_{\delta}} \simeq $ 0.496. This value is close to 
$T_{U_{\xi}} \simeq $ 0.471, estimated from the high temperature 
divergence of $\xi(T)$. Distinct values of  
$T_{U_{\delta}}$ and $T_{c}$ point clearly to the presence of two
 transitions in this system. In Fig.~\ref{fig:7} we show the temperature 
 variation of $\textit{D}_\textit{t}$ along with its temperature derivative at 
\textit{L} = 150. $\textit{D}_\textit{t}$ shows a rather sharp change in 
its slope at the location of the $C_{v}$ peak, (which is absent in the 
\textit{B}-ensemble data), and the profile of its temperature derivative is
qualitatively very similar to that of $C_{v}$ (Fig.~\ref{fig:1}). The 
source of energy contribution to the observed $C_{v}$ peak is closely 
connected with the rate of the defect density production, - a clear 
manifestation of the qualitative changes in the relevant length scale 
in this temperature region.    

The crossover of the system as it is cooled initially from a high 
temperature disordered phase is connected with the development of an 
additional relevant scaling field in the parameter space, as the 
increasing correlation length (arising from the diminishing number 
density of the unbound defects) matches typical nearest neighbour 
distances in lattice units. The apolar nature of the local directors 
associated with the lattice sites ($Z_{2}$ symmetry) in the presence 
of an attractive biquadratic Hamiltonian, and facilitated by the extra 
degree of molecular reorientation (\textit{n} = 3), lead to the 
development of critical nematic clusters at these coherence lengths. The 
transition at $T_{c}$ is thus from the isotropic to a nematic phase, 
with the $\xi(T)$ on either side of the transition at $T_{c}$ being 
limited by different mechanisms, -  critical fluctuations on the high 
temperature side and nematic director fluctuations at low temperatures -, 
thus forbidding the development of QLRO regime. The system crosses over 
towards this fixed point for \textit{T} $\leq$ 0.6. The phase immediately 
below the $T_{c}$ with a single temperature-dependent length scale 
represents an interesting nematic medium which has unbound topological 
defects undergoing thermally activated kinetic processes, but their length 
scales being limited by the coherence lengths of the director fluctuations. 
Similar scenario was noticed earlier in fully frustrated anti-ferromagnetic
 Heisenberg (HAFT) model on a triangular lattice which also hosts $RP^{2}$ 
topology for its order parameter space \cite{Kawamura}. In this 
(\textit{d} = 2 and \textit{n} = 3) magnetic model, spin fluctuations 
were found to limit the diverging correlation length in the high temperature
paramagnetic phase. The second transition in the LC system estimated by 
$\delta$ at $T_{U_{\delta}}$ ($\simeq$0.496), is mediated by the 
topological defects, leading to a lower temperature nematic phase with 
growing average cluster size and hosting only bound topological defects. 
In HAFT models, such media were referred to as spin gel phases. 
 
We conclude that the presence of the extra dimension for molecular freedom, 
coupled with the local gauze invariance under $Z_{2}$ symmetry are ideally 
suited to induce a crossover when the development of the orientational 
correlation length to a sufficient degree makes the attractive interaction 
of the LL model a relevant scaling field. The failure of the earlier MC 
studies to associate a size-invariant Binder's cumulant at the proposed 
topological transition temperature seems to be justifiably conjectured as 
due to an underlying crossover, but the conventional MC sampling method 
could not access the higher energy microstates apparently needed in this 
temperature region (Inset of Fig.~\ref{fig:5}) to locate the true minimum 
free energy configurations, thus missing to detect the crossover phenomenon.  
 
We acknowledge the computational support from the Centre for  Modelling
 Simulation and Design (CMSD) and the School of Computer and 
Information Sciences (DST PURSE - II Grant) at the University of
 Hyderabad. BKL acknowledges financial support from Department of
 Science and Technology, Government of India  vide grant ref No: 
 SR/WOS-A/PM-2/2016 (WSS) to carry out this work.

\end{document}